\documentclass[5p]{elsarticle}

\usepackage{amssymb}
\usepackage[hyphens]{url}
\usepackage[hidelinks]{hyperref}
\usepackage[capitalise]{cleveref}
\crefname{appendix}{}{}
\usepackage{colortbl}%
\usepackage{xcolor}%
\usepackage{booktabs}
\usepackage[draft]{fixme}
\journal{Computer Communications}
\hypersetup{breaklinks=true}
\title{}

\title{Do we need a Contact Tracing App?}

\author[1]{Leonardo Maccari}

\author[2]{Valeria Cagno}

\address[1]{Department of Environmental Science, Informatics and Statistics \\ Universiy of Venice}
\address[2]{Department of Microbiology and Molecular Medicine, \\University of Geneva Medical School}
\begin{document}
\begin{abstract}
The goal of this paper is to shed some light on the usefulness of a contact tracing smartphone app for the containment of the COVID-19 pandemic. We review the basics of contact tracing during the spread of a virus, we contextualize the numbers to the case of COVID-19 and we analyse the state of the art for proximity detection using Bluetooth Low Energy. Our contribution is to assess if there is scientific evidence of the benefit of a contact tracing app in slowing down the spread of the virus using present technologies.  Our conclusion is that such evidence is lacking, and we should re-think the introduction of such a privacy-invasive measure. 
\end{abstract}


\maketitle

\section{Introduction}

During the recent pandemic of the coronavirus infectious disease 2019 (COVID-19) national governments imposed various flavors of ``lockdown'' forcing people in their houses and preventing many of them to work, with the goal of slowing down the pandemic and make it more manageable by national health systems. 
Such measures are unprecedented as they limit personal freedom and strongly impact national economies and need to be maintained only for the shortest time required to stop the emergency. 
When the emergency is tamed, governments introduce a so-called ``phase 2'' which imposes milder limitations that can progressively help economies recover.
Yet, being the first time in modern history that our society faces such a threat, we don't have a clear path to follow for phase 2. On the one hand, there is a widespread interest in removing personal limitations, on the other, the risk of provoking a second wave of spread of the virus will be present until a vaccine or a cure are found and made available to everybody.

One of the measures considered by many governments, and implemented by some, is the introduction of a mobile phone application that performs contact tracing, e.g., it provides a list of contacts that took place between couples of people in a certain time span. The app enables a person that was tested positive to Sars-CoV2 (the aetiological agent of COVID-19) to send a warning to all the contacts. 
Those who receive the warning will presumably act accordingly (isolating themselves, or asking to be tested) and this is imagined to contribute to keep the diffusion of the virus under control. 

In a few weeks the interest for this application skyrocketed, Google and Apple produced a dedicated API to support contact tracing, the European Union provided guidelines to specify the privacy implication of such applications and a consortium of companies started to elaborate proposals. The public debate rightly focused on the privacy aspects, which are of paramount importance because this is, in essence, \emph{the first ever mass-scale contact tracing action promoted by democratic states}. The privacy fallout caused by the extensive use of such an app may be huge and the analysis of the correlated risks was the subject of several works \cite{bock2020data,cho2020contact,troncoso2020decentralized}. If our society takes such risks, we must expect a greater collective payoff. 

Unfortunately, little importance has been given to the effectiveness of this application in limiting the diffusion of the virus. 
Here we take a first principle approach, and review the research literature to understand if the current technical means may correctly perform contact tracing as needed to stop the pandemic. 
If this is not true, then the whole debate about the privacy implications may simply lose ground. Our approach does not rely on the analysis of each single application, as there are other ongoing works that cover their details \cite{ahmed2020survey,cunche2020using}. Instead, we use an interdisciplinary approach:  we analyze some real world constraints on what a contact tracing app may be allowed to do, we interpret the data coming from the medical literature about the dynamics of viral spread, and we review the technical literature to understand what technology provides.

In our analysis we were not able to find any strong evidence that a privacy preserving app based on current technologies can achieve the stated goal. We observe that with current technologies a similar app could be used to perform a mild risk assessment, but this approach needs more technical discussion and social acceptance. Therefore, the adoption of contact tracing apps has to be rethought. 

In the rest of the paper we discuss what is needed to slow down the spread of COVID-19 (\cref{sec:monitoring}), and what are the privacy and technical constraints we have to respect for the app to be effective in a short time (\cref{sec:tracingmobile,sec:proximitydetection}). We then review the experiments carried out in the literature in similar realistic conditions (\cref{sec:soa}). All these elements constitute the basis for a grounded discussion on the use of a contact tracing app (\cref{sec:discussion,sec:wayahead}).

\section{Monitoring the Contagion}
\label{sec:monitoring}

The transmission of Sars-CoV2 primarily happens due to respiratory droplets \cite{who2020report,liu2020community}, therefore reducing the number of contacts of an infected person is a key factor to slow down the contagion. This is achieved with social distancing, hygiene measures and isolation of infected people. It is intuitive that the earliest an infected person is isolated, the lowest are the chances to have contacts and spread the virus. Therefore, the rationale for contact tracing is straightforward, when a person is tested positive, we need to quickly identify those that were potentially infected by this person so that we can isolate them as well. For COVID-19 there are evidences that the transmission could occur also from asymptomatic and pre-symptomatic people \cite{he2020temporal}, these people may not even known to be infected, which makes it even more urgent to quickly isolate the close contacts of a person that was tested positive.

The key parameter to monitor the speed of the diffusion of contagion is the reproduction number $R$, which expresses the expected number of people infected by one single individual at a certain stage of the epidemic. $R$ depends on a number of factors, including the way the virus is transmitted but also the contention measures that are enforced. The higher is the value of $R$, the faster the virus spreads. The goal of contention measures is to lower $R$ below 1, so that the total number of infected people decreases with time. The base reproduction number $R_0$ is the value of $R$ at the beginning of the contagion, when all the population is susceptible of being infected and no contention measures are in place, thus, it generally holds that $R \leq R_0$.

There are estimates of the value of $R_0$ using data that refer to initial phases of the outbreak when no contention measures where in place and all the population was susceptible. 
One of the first attempts to estimate $R_0$ was performed observing the passengers of the Diamond Princess cruise ship \cite{zhang2020estimation} and reports a median value of $R_0 = 2.28$. A recent modeling analysis of the initial phases of the outbreak in China estimates a $R_0 = 3.54$  \cite{zhao2020preliminary}.
When contention measures are put in place $R$ is expected to decrease, for instance at the time of writing the value of $R$ in Germany is estimated to be lower than 1 \cite{koch2020report}, as well as in Italy in the period between April 19th and May 7th, and in Wuhan (where the contagion started) the number estimated after the lockdown measures was 0.28 \cite{iss2020epidemia}.

Of course the average or median estimated value of $R$ may not be representative of some extreme cases, the presence of \emph{super-spreaders} has been reported for COVID-19 \cite{liu2020secondary} with individuals possibly infecting up to 10 other people in a single event or in extreme cases up to 32. Yet, this kind of event is likely at the early stage of the epidemic but can be controlled with contention measures that have a limited impact on personal freedom: maintaining interpersonal distance, using face masks and hygiene measures, and forbidding large in-person meetings or taking additional measures such as lists of participants in those cases. In the phase 2, these measures must be already in place.

\subsection{The Secondary Attack Rate and the Contact Type}

Another element that is relevant for our analysis is the so-called secondary attack rate (SAR), that is, the percentage of people that is actually infected among the contacts of a person that has been tested positive to Sars-CoV2.  In the early stages of the outbreak of the virus in China (Jan 14th - Feb 15th, 2020) a group of people that tested positively were observed together with 1286 people among their contacts \cite{bi2020epidemiology}\footnote{Contacts were identified as ``\emph{those who lived in the same apartment, shared a meal, travelled, or socially interacted with an index case}'' and excluded ``\emph{contacts (eg, other clinic patients) and some close contacts (eg, nurses) who wore a mask during exposure}''.}. 
The estimated value of SAR was 6.6\% (81 over the 1286), while in more recent works that analyze the evolution of th Pandemic in Taiwan, SAR was estimated to be 0.8\% (32 out of 3795 \cite{huang2020estimation}).

The secondary attack rate can be further specified for different contacts types (family, workplace etc. \ldots). Among the contacts in the Chinese study, the large majority of those that tested positive (77 over a total of 81 positives) were households of the infected person. 
A similar statistics related to the advanced phases of the contagion comes from the Istituto Superiore per la Sanità (ISS, the main center for research, control and technical scientific advice on public health in Italy) which provides weekly updates on the evolution of the Italian situation. The report referring to April 7th - May 7th \cite{iss2020epidemia} includes the distribution of the contact places for more than 9,360 cases\footnote{In Italy the contention measures have been adopted with three steps of increasing intensity in March 1st, 9th and 22nd, were reduced starting May 4th and were still in place in the considered period.}. \Cref{tab:iss} summarises the data and shows several interesting facts, the first is that in Italy 58\% of the new infections in the observed period happened in retirement homes. This is due to the management of the emergency in Italy, which made retirement homes a focus of infection. Albeit this happened in other nations, it can not be fully generalized. For this reason in the last two columns of the table we report the distribution of cases excluding the first line. 
\begin{table*}
	\begin{center}
	\begin{tabular}{llllll} 
		\toprule
		Location & Number of infected & Percent & Cumulative &  Percent w/o RH & Cumulative w/o RH\\ 
		\midrule
		\rowcolor{gray!10} Retirement home (RH) &5,468 & 58.4 & 58.4 & - & - \\
		\rowcolor{gray!10} Family & 1,712 & 18.3 & 76.7 & 44 & 44 \\
		\rowcolor{gray!10} Hospital/clinic & 816 & 8.7 & 85.4 & 21 & 65\\
		\rowcolor{gray!10} Workplace & 228 & 2.4 & 87.9 & 5.9 & 70.8 \\
		\midrule
		Boat/Cruise  & 83 & 0.9  & 88.8 & 2.1 & 72.9 \\
		Religious Community & 64 & 0.7 & 89.4 & 1.6 & 74.6  \\
		Other & 989 & 10.6 & 100 & 25.4 & 100\\
		\bottomrule
	\end{tabular}
	\caption{The location of contagions in Italy, April 7th - May 7th. Column 5 and 6 report the same data excluding the numbers related to Retirement homes.}
	\label{tab:iss}
	\end{center}
\end{table*}
We can see that about 87\% (70\% excluding retirement houses) of the cases happen in extremely predictable locations: 18\% (44\%) in the family, 8\% (21\%) in hospitals, and 2.4\%  (5.9\%) in workplaces. Only 12\% (29\%) of the cases happen somewhere else.

\subsection{The Needle in the Haystack}
\label{sec:needle}

\begin{figure}
 	\centering	\includegraphics[width=0.85\columnwidth]{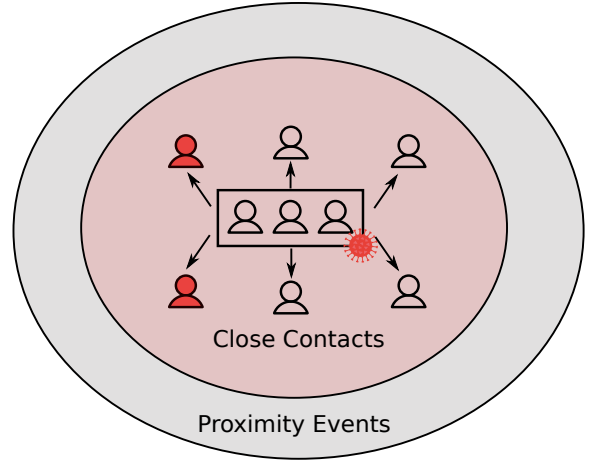}
 	\caption{A schema of contact tracing, every three infected people, with $R=2$ we are looking for 6 infected more people, among which two of them are expected not to be in the group of close contacts. The close contacts are part of the group of those considered \textit{in proximity} of the infected people.}
 	\label{fig:groups}
 \end{figure}

Previous sections exposed data generally overlooked when discussing about contact tracing applications: i) every infected person infects in average 3 people in the early stage of the epidemic, a number that decreases when contention measures are applied; ii) the large majority of the infections involves extremely predictable groups of people, family members and work mates which can be identified with basic means, i.e. with manual contact tracing among households and colleagues.
If we consider a pessimistic value $R=2$ for the phase 2, and we assume that 29\% of the infected people can not be traced with manual contact tracing (we derive this figure from the data in \cref{tab:iss}), then we have an average rounded value of approximately 2 non-identified infected contact per 3 infected people. 
Moreover, we observed that SAR for COVID-19 is low: only a small percentage of the close contacts monitored by medical authorities were tested positive, which means that the definition of close contact is already coarse enough to produce a large number of false positives.
\Cref{fig:groups} summarizes the situation, with the numbers discussed so far: from a group of three infected people we expect 6 more infected people ($R=2$), since the SAR is between 1\% and 6\%, then the close contact group is in the orders of hundreds of people. Among them only two out of six will not be easily traced with analog contact tracing. The set of close contacts is included in the set of people that were in the proximity of the infected people (which we discuss in the next section).

The requirements that emerge from the medical literature suggest that digital contact tracing, to be useful must be extremely precise: we need to identify a small number of people that are not easily tracked with analog contact tracing (two every three infected people) and we can not arbitrarily enlarge the amount of identified close contacts, because the percentage of infected people among them is very small.

\section{Contact Tracing with Mobile Phones: Motivations and Privacy Guidelines}

\label{sec:tracingmobile}
Contact tracing refers to identifying ``close contacts'', a term that is specifically defined for a certain virus. For COVID-19 the World Health Organization defines a ``close contact'' as ``\emph{Any person who had contact (within 1 metre) with a confirmed case during their symptomatic period, including 4 days before symptom onset}''\footnote{See \url{https://www.who.int/publications-detail/the-first-few-x-(ffx)-cases-and-contact-investigation-protocol-for-2019-novel-coronavirus-(2019-ncov)-infection}}. 
This definition leaves room for interpretation because the duration of the period is uncertain and does not take into consideration prevention measures that could be adopted (such as face masks). Several national health systems provide a  more specific version, the ones we were able to review generally consider close contact as being at short distance (less than 2 meters) with an infected person for more than 15 minutes (see \cref{app:contact}).
One of the key challenges of analog contact tracing is that the person that is positive to tests needs to be interviewed by an expert that is able to identify the close contacts, based on the kind of exposure. This is a time-consuming task, and people can not always remember all their contacts. The rationale of a contact tracing app is that we can identify new contacts that for some reason did not emerge from the interview, and that we could speed up the notification to the close contacts with notifications on the smartphone.

Mobile phones can not be used for contact tracing but can be used for ``proximity detection'', which means, in a nutshell, identifying couples of devices that have been in communication range with each other for a specific time. At the base of the use of mobile apps lies the assumption that the set of people for which a proximity event was recorded is a good approximation of close contacts.

One of the works that is often mentioned in support of a contact tracing application is from Ferretti et al. \cite{ferretti2020quantifying}. 
Ferretti et. al model the outbreak of COVID-19 using the state of the art data on its diffusion and show that considering the specific characteristics of COVID-19, a fast contact tracing can be beneficial to lower the value of $R$ below 1. The authors assume that a contact tracing app can help speeding up this task, but they do not enter into the details of the performance of the application. They just consider the app as a factor that can speed up the detection of new infected people. In this sense, it is quite straightforward that any factor that can make detection faster will reduce $R$, and the paper provides a quantitative analysis of this reduction using the efficacy of contact tracing as one of the parameters of the model. 

On this basis, several governments started to express interest on the adoption of such an app, which of course initiated a debate on its privacy implications. It is clear that collecting data on contact tracing introduces a privacy risk, which must be mitigated using a proper app design. This sparkled an intense public discussion on several key themes such as the app model (centralized/distributed), on the adoption (opt-out/opt-in), and on the data to be collected (only proximity or any other relevant data, such as position using GPS). 
A deep analysis of the privacy and security implications of contact tracing apps is out of the scope of this paper, and can be found in other works that were published in the months following the outbreak of the pandemic \cite{cho2020contact,bock2020data,troncoso2020decentralized,ahmed2020survey}. In the context of our paper we are interested in the outcome of this debate, which can be summarized in two landmark events.
The first was the publication from the European Union of guidelines for Privacy-Preserving contact tracing apps. These guidelines must be respected for apps to be acceptable under a privacy point of view in Europe. The second was the publication by Apple and Google of an API for \emph{Exposure Notification} (EN), the building block required to perform proximity detection on their mobile operating systems. Such API is compatible with the European guidelines. These events set the stage for the development of any contact tracing app, because having the support from vendors covering almost 100\% of the mobile phone market is essential to reach a large adoption, and respect of privacy guidelines (at least in the European panorama) is essential. 
In fact, recent works analysing some of the available contact tracing apps show that in the continental Europe among the 8 monitored apps 5 use the broadcast model (essentially the EN model we describe later on), while 3 use the broadcast model and rely to a connection-based model when the first is not available \cite{cunche2020using}.
Therefore, without entering into the privacy debate, the goal of this paper is to address the following question: \emph{do we have enough evidences suggesting that the proposed proximity detection solution can achieve the necessary accuracy in contact tracing while respecting the privacy guidelines?}

To answer this question we need to briefly review the technical guidelines provided by the EU, and the technology proposed by Apple and Google. 

\subsection{Contact Tracing: the EU Privacy Technical Guidelines}

The people we meet every day and the duration of the encounters are extremely sensitive information. For this reason, the European Union published guidelines on the use of contact tracing in the context of the COVID-19 pandemic \cite{eu2020guidelines} which we take as functional requirements for contact tracing apps. Among them, some of the points that are relevant for our analysis can be summarized as follows (\cref{app:docs} reports excepts from the mentioned document that expand the bullet points):

\begin{itemize}
	\item Contact tracing must be based on voluntary adoption, there should be no consequences for those that opt-out
	\item Phone location should not be used, only proximity data should be used
	\item It should not be possible to track back the identity of a person using the data from the app. This is a crucial point, when Alice receives the notification that she was in contact with an infected person, she should not be able to say if this person is Bob\footnote{We use the widespread cryptography jargon describing two users of the app as Alice and Bob and the server that is their intermediary as Eve.}. 
	\item Information should not exit the user phone if not absolutely necessary.
\end{itemize}

These guidelines go in the direction of retaining the least possible information and maintaining it as much as possible in the user device, and not in some centralized server. 

Another extremely important guideline states the importance of false positives. People that are diagnosed with COVID-19 will be subject to isolation and thus decisions can not be taken with an automated mechanism \cite{bock2020data}. 
Beyond personal consequences, false positives may have two side effects that can make the app useless, or even detrimental. If the rate of false positives is too high, people receiving alerts will simply start to ignore them, thus defeating the goal of the app itself.  If instead people do not underestimate the alerts and receive many false positives, huge testing capacity, large enough medical staff to perform the tests, and efficient logistic to avoid risk of transmission in the hospital setting will be required. All these resources are scarce during the upsurge of an epidemic and can not be wasted due to the inaccuracy of contact tracing.



\section{Proximity Detection with Bluetooth Low Energy (BLE)}
\label{sec:proximitydetection}

BLE is a natural candidate to perform proximity detection between pairs of mobile phones. It is a well established technology, introduced in 2010 and currently part of the core Bluetooth specification\footnote{See \url{https://www.bluetooth.com/blog/bluetooth-low-energy-it-starts-with-advertising/}.}.
There are no publicly available statistics for the market uptake of BLE in active devices at the time of writing. According to Katevas et al. \cite{katevas2016detecting} BLE is present in almost all the existing iPhones, and the Bluetooth SIG estimates that in 2024 100\% of new devices will be equipped with BLE\footnote{See \url{https://www.bluetooth.com/bluetooth-resources/2020-bmu/}.}. Support for BLE is available in Android since version 4.3 (2013) and in iOS since version 5 (2011), so it is reasonable to assume that a very large portion of the mobile phones on the market support BLE.

Proximity detection can be performed with BLE with a simple mechanism. Each BLE-equipped device can be in two states, the broadcaster or the observer. The broadcaster sends a broadcast beacon message on three default channels every ``Advertising Interval'', the observer instead every ``Scan Interval'' wakes up and listens to beacons for a ``Scan Window'' time. 
When the observer receives the beacon it estimates the distance from the broadcaster using the Received Signal Strength Indication (RSSI). We focus only on the description of the solution that Google and Apple provided to perform Contact Tracing using BLE, while generic descriptions of BLE can be easily found in the literature \cite{montanari2019devising}.

\subsection{Google/Apple Privacy-Preserving Contact Tracing}

In April 2020 Apple and Google released a joint document with the technical specifications of a Privacy-Preserving Contact Tracing API supported by their operating systems\footnote{Privacy-Preserving Contact Tracing, see \url{https://www.apple.com/covid19/contacttracing/}.} based on BLE. The specification is called Exposure Notification and offers a trade-off between energy consumption, user privacy, and efficacy. It is supposed to become the layer on which every contact tracing application may be based on. Nothing prevents developers to use other technologies, but considering the large diversity of devices and OS versions in the market it is unlikely that any custom solution may reach the needed uptake to be effective. 

When using EN, a phone acts both as a broadcaster and as an observer. The Advertising Interval is set between 200 and 270 ms (corresponding to approximately 4 Hz), while the Scan Interval is not specified, it should ``\emph{have sufficient coverage to discover nearby EN Service advertisements within 5 minutes [\ldots] with minimum periodic sampling every 5 minutes}''. In the API specification it is mentioned that scanning happens typically every five minutes, with a Scan Window of 4 seconds\footnote{See \url{https://developers.google.com/android/exposure-notifications/ble-attenuation-overview\#aggregation\_over\_a\_scan\_scaninstance\_min\_and\_average}}. 
EN defines for each device a Temporary Exposure Key (TEK), which changes once per day. This key is kept in Alice's phone and never leaves it if Alice does not get infected. Every ten minutes this key is used together with a counter ranging from 0 to 144 to generate another key, the Rolling Proximity Identifier (RPI). The RPI is included in the beacon (together with some meta information that are not relevant for this discussion).
Alice's phone is thus identified by the same RPI for 10 minutes, which will give Bob the chance of being an observer for two Scan Windows. Without packet loss Bob will receive in average 32 beacons to estimate the distance with Alice's identifier before it rotates. Bob stores all the RPI he receives in his device.

If Alice at some point becomes infected she uploads all the TEK for the last 14 days (or any other value decided by the application) to a Diagnosis Server (Eve), that is run by the app provider (not Apple or Google necessarily but the developer of the app). Eve periodically aggregates the keys ``\emph{from all users who have tested positive, and distributes them to all the user clients that are participating in EN}''.
Bob then periodically receives sets of TEK keys coming from many people tested positive to COVID-19, he re-generates all the RPIs and checks if some of them are present in his own local storage. If some of the keys are present in his storage, he was in the proximity of an infected person.

The rationale of EN seems to be following:
\begin{itemize}
	\item minimize the energy consumption. Since users are expected to constantly run this system, it should not severely impact battery use.
	\item minimize the amount of exchanged information between Alice and Bob. There is no unicast communication happening or packet handshake. This also makes power consumption predictable as it does not depend on the number of devices nearby.
	\item minimize the amount of information transmitted to Eve. Eve does not store the contacts of Alice and Bob.
	\item provide to Bob a sufficient amount of information to estimate his exposure to some infected person without revealing who these people were.
\end{itemize}

Note that Apple and Google do not take responsibility for deciding when to notify Bob, the task of passing from Proximity Detection to the definition of Close Contact is completely delegated to the app.

\subsection{Privacy of EN}

EN seems to be aligned with the privacy guidelines from the EU. In this regard it is worth mentioning the notable work done by the DP-3T consortium, a group of international experts that are devising distributed, privacy-aware solutions for contact tracing. In a white paper they describe three solutions, two distributed ones and a centralized one, and analyze their privacy characteristics \cite{troncoso2020decentralized}. One of the distributed solution is very close to the EN proposal and their analysis basically confirms that EN can be used minimizing the risk of privacy breach. 

Yet, minimizing risk does not mean removing it. A deep discussion on the privacy risks related to contact tracing are out of the scope of this paper, we just mention two issues, as concrete examples. 
First is that when Alice is tested positive, if Bob and Alice spent time in proximity, Bob should receive enough information to know that there was a close contact with an infected person, but not enough to understand who this person was. Alice has the right of not letting others (beyond medical staff) know she is infected. In practice this is impossible to guarantee with a decentralized solution. Bob will receive all the RPIs Alice generated and he can associate to each RPI a short time-interval in which he was in proximity of an infected person. If the contact lasted for a long enough time there are chances that Bob may infer that Alice was, for instance, the person that was sitting next to him in the office for all the covered period. 
Second, is that even in the decentralized model, there is a server that receives from the apps the notification that the app owner was tested positive. The server does not need to store the association between the app owner (from which the message is coming) and his/her health state, but there is still one point in which this information could potentially be collected. This opens the way to attacks to the servers from criminals that may be interested in collecting such information. 

As risks exists and their impact can be amplified by the massive scale of the adoption of contact tracing applications, we need to properly assess the benefit we expect this application could produce.

%
%
%
%

\section{Proximity Detection with BLE: State of the Art}
\label{sec:soa}
In this section we review the works in the literature that deal with proximity detection using BLE on mobile phones. Indoor localization has been a hot research topic in the last few years and the estimation of the distance between devices is the building block of any localization system. Our intention was to restrict the analysis to those works that provide insights on the applicability of BLE contact tracing for the COVID-19 use case, i.e. those that respect the following criteria: 
\begin{enumerate}
	\item The work must present a real implementation.
	\item The experimentation must involve commercial off the shelf mobile devices, and not only custom devices.
	\item Proximity detection should be performed without external aids if not for results validation. This excludes fixed BLE beacons or implicit constraints due to the set-up of the experiment (e.g., the experiment takes place in a single room only).
	\item The proposal must be compatible with EN: it must use BLE, it should not require post-processing of data by a centralized entity and should not require handshakes between devices.
\end{enumerate}

Unfortunately, none of the works in the literature satisfied these criteria. We decided then to analyze two sets of works, the first one provides an overview of the challenges in the distance estimation using BLE, which is itself a non-trivial task. The second set describes experiments that are as close as possible to the use case of contact tracing for COVID-19, in order to assess what are the main challenges for a real contact tracing app. 

%
%
%

\subsection{Estimating Distance with BLE}

Indoor positioning is a hot research topic and there are many works in the literature that deal with it (see Gu et al. for a recent literature review \cite{gu2019indoor}). In most cases indoor positioning is obtained with static beacons (i.e. dedicated devices that behave as BLE broadcasters). The mobile devices receive the messages from the beacons, measure the RSSI and use this information to estimate their position. Even if this approach is not usable for contact tracing, the works in the literature can provide insights on the performance of distance estimation. EN uses RSSI with BLE and thus it is important to shed some light on its accuracy, even only in controlled environments.

Given the RSSI from a transmitting device, distance can be computed as:

\begin{equation}
D = 10^{\frac{C_0 - RSSI}{10n}}
\label{eq:d}
\end{equation}
where $D$ is the estimated distance, RSSI is the received signal strength, $C_0$ is the average RSSI value at a reference distance from the transmitting device (typically 1m),  and $n$ is the decay exponent. The values of $n$ and $C_0$ are not universally given, they depend on the transmitter and on the environment, and they need to be estimated for each device and for each environment. As an example, Mackey et. al \cite{mackey2020improving} make an estimation of $C_0$ for three beacon devices whose specification stated the same value of $C_0$, and measure a difference up to 10dBs. The EN documentation does not enter into the details of distance estimation. EN provides only an estimation of the attenuation of the signal over its path, and leaves to the implementer the choice for the thresholds to identify a close contact. EN provides a list of transmission power and correction factors for a number of devices that were tested, but still, the parameters of \cref{eq:d} are in general uncertain. 

Even assuming that the parameters of \cref{eq:d} are fixed, the second source of error is the estimation of the RSSI. If we set $C_0 = -82.42, n = 1.96$ (extracted from \cite{mackey2020improving}) we see that an error of 1 dB at a distance of roughly 1.5m\footnote{The error is non linear, we use 1.5m as a reference as it falls between 1m and 2m, which are the cut-off distance for contact tracing adopted by WHO and several countries.} produces an error of roughly 20cm. 
Which are the possible causes of error? First of all, the value of RSSI provided by the devices is not a standard measure. As noted in the literature \cite{kaczmarek2016accuracy} the value received by an application is the result of the elaboration made by hardware, drivers and software, and there is no standard expectation on its accuracy. Second is multipath fading: the RSSI is the sum of the signal that is directly received in the line of sight and/or the reflections from the objects in the surrounding environment, which changes constantly. Third, BLE broadcasts beacons on three different Bluetooth channels and the response of the smartphone radio in the three channels is different, which provides a very noisy figure when the levels are summed into a single value. Finally, even the orientation of the phone influences the RSSI, so that the same couple of devices, at the same distance may measure a different RSSI due to phone orientation. Considered altogether, these factors make the the parameters of \cref{eq:d} extremely variable depending on the environment, the devices, the position of the phones, and thus distance estimation using RSSI is simply noisy in generic conditions. 

If we look at the results in the literature we can try to quantify the expected error\footnote{Unfortunately, the absence of open data does not enable us to extract exact figures from the works in the literature. We need to rely on visual inspection of graphs and therefore, our observation are quantitative but can not be easily systematized. Still, the inaccuracy we outline is so macroscopic that justifies our reasoning.}. 
Katevas et al. \cite{katevas2016detecting} performed detailed experiments to estimate the accuracy of distance estimation with BLE in a very controlled environment, including an anechoic chamber. The results show that distance estimation on commercial devices (iPhone 5S and 6S) are very noisy, with en error around 0.75m at the 1.5m distance, and 1.5m on an 3m distance. Neburka et al. \cite{neburka2016study} provide another interesting insight on the behaviour of BLE, showing that the same device (an RN4020 BLE module, not a mobile phone) when using different channels (again, in an anechoic chamber) can show up to 15 dB variation in RSSI measurement, depending on the channel. 

Several papers \cite{faragher2015location,zhuang2016smartphone} perform an analysis of the accuracy of a BLE-based positioning system when using mobile phones as receivers and fixed, dedicated devices as broadcaster. Again, they measure RSSI variations of tens of dBs, for instance, when people walk there can be drops in the measured RSSI up to 30 dB \cite{faragher2015location}. Similar measures are performed also by Montanari  \cite{montanari2019devising}, and the interesting observation that the author does is that the RSSI value is different from the ones measured in other works \cite{faragher2015location}, as a confirmation that even in controlled environments, the measured values differ largely.
Naghdi and O'Keefe quantify the effect of human bodies that shadow the propagation of BLE signals \cite{naghdi2020detecting} and show they can produce a drop of tens of dBs in the RSSI values, with a large variability due to distance from the source and the channel used. 

Finally, the EN library documentation provides a procedure to calibrate distance estimation. The documentation reports that RSSI measured using two phones in an anechoic chamber varies up to 10 dBs just for a change in the orientation of the phone, without changing any other environmental parameter\footnote{This value is reported by Google in the description of the procedure necessary to calibrate the parameters of EN, see  \url{https://developers.google.com/android/exposure-notifications/ble-attenuation-procedure\#device-orientation}}.

Summing up, the literature provides enough evidence for the fact that even in favorable conditions (anechoic chamber, controlled scenario, known hardware for the transmitter) several sources of noise will affect the received RSSI up to \emph{tens of dBs}, which translates in distance estimations error up to several meters. While this can be tolerable in certain applications (for instance the location of a person inside a room) it completely defeats the goals of contact tracing that should identify contacts in the range of 1-2 meters. 

\subsubsection{Improving the Estimation}

 Mackey et al. \cite{mackey2020improving} introduce several filtering techniques and test them to estimate the distance of a phone from a beacon. 	
They achieve encouraging results, with a mean absolute error for distance estimation (below 3m) given by 0.27m and 0.412m (in large or small room) for the best chosen filtering technique. Unfortunately, these results are far from reproducible in the contact tracing environment: they use static beacons as transmitters, in line of sight with the phone, they require training (4 out of 5 filtering techniques use internal parameters that were calibrated on the scenario, with fixed thresholds), the scenario is completely controlled (empty room, no people), and before beginning the experiment, 1000 data points for each distance are collected to estimate $n$ and $C_0$, with 1200 more data points used to make the estimation (against the 32 we expect to have with EN).

To improve the distance estimation is possible to use BLE together with data coming from other sensors on the mobile phone. Liu et al. \cite{liu2020real} use the phone accelerator and gyroscope to improve localization, inferring the walking direction of the user.  Alas, also phone sensors are imprecise and they are sensitive to the way the user holds the phone therefore this approach requires calibration for each specific device. Multi-modal positioning has been proposed using a mix of BLE, Wi-Fi, and visual landmark recognition \cite{dumbgen2019multi}, which requires the user to periodically re-calibrate the app in known positions. These recent papers show once more that  distance estimation based only on BLE is extremely noisy in real world settings, and describe promising research direction that will be explored in the future. For the time being, due to their complexity and their requirements, they have been shown to work in a controlled environment, but they are far from being usable for the COVID-19 contact tracing.

\subsection{``Contact Tracing'' with BLE}

Montanari et al. \cite{montanari2017study,montanari2019devising} use BLE to perform contact tracing in an office environment, with a set-up that is the most similar to a real world situation. The goal of the experiment was to track interactions among 25 co-workers in an office for 4 weeks, data were validated by human observers in the office and stationary beacons. The ground truth consisted in 401 observed interactions, meaning two or more people standing at less than 3 meters from each other and having a conversation. On average the interactions lasted for 1 minute and 13 seconds and 70\% of the interactions were shorter than 1
minute, while only 5\% were longer than 5 minutes. The authors use a custom device in order to achieve a high precision in data collection, but they then re-sample data in order to match the configuration that is achievable on commercial devices. Some of the results are encouraging, with realistic configurations that could achieve between 81\% and 96\% true positive detection rate.

Unfortunately the experiment set-up is far from the COVID-19 use case as it breaks all the criteria we defined. Devices were smart watches and not smartphones, this makes a big difference because watches are always at people's wrist and can not simply be left on a desk. With smart watches there are higher chances that the two devices are in line of sight, while a smartphone usually stays in a pocket or bag, and RSSI strongly depends on shadowing. The choice of the parameters allowed a much more fine-grained sensing than\ EN, as Scan Interval and Scan Window were set to be below 5 seconds, orders of magnitude lower than with EN. The testing environment was a firm office and only that office, participants were asked to remove watches when going out of the office. Each contact was tracked using the RSSI measured on both watches in order to mitigate the effects of multipath fading. To achieve this, all data were stored in a server and later on post-processed. Post processing used machine learning to identify contacts, with a supervised learning approach. All these issues make the set-up not at all comparable with the COVID-19 use case, and make it impossible to generalize the results.

Girolami et. al. \cite{girolami2018detecting} investigate the possibility of using smartphones for contact tracing. In this case the experiment included students from a high-school that were asked to perform certain interactions (such as standing or sitting in front of each other for 3 minutes) while their mobile was recording BLE messages. The reported accuracy of encounter detection reached almost 82\%. 
Unfortunately, again the testing conditions were not comparable to the COVID-19 use case. Interactions were not spontaneous, the participants were asked to perform specific actions, and these actions were happening in certain places, not ``in the wild''. The whole data-set was collected and post-processed, the reported accuracy was obtained with the best combination of tracking parameters and considering received beacons on both mobiles involved in the contact. A key contribution to this discussion from Girolami's work is contained in this sentence: \emph{Firstly, we investigate the possibility of using commercial smartphones to advertise and to collect BLE beacons demonstrating that, currently, such approach is not feasible due to the heterogeneous implementation of BLE firmware in different versions of mobile OS (both Android and iOS)}. The authors initially tried to use the devices of the students but found out that in the batch of Android devices owned by the participants (which matched 15 different models), 42\% of them were not usable for contact tracing. Even if the hardware and the software were supposed to be compatible with BLE, the device simply did not allow BLE to be used for active beaconing. In the end, the authors used mobile phones as observers but had to equip participants with a BLE watch acting as the broadcaster. 
Even once the broadcaster was set to be a ``standard'' watch with fixed hardware and software features, and the receiver phone was kept in a pre-defined position (front pocket or back pocket in participant's pants), the measured RSSI changed substantially depending on the mobile phone receiving the data and its position. Furthermore, the median number of lost beacons per session was larger than 50\%, which suggests that in a real world scenario the loss of beacons is an extremely important factor, which can be mitigated only with a large Scan Window, which is known to be the highest source of power consumption.
The trade-off between battery consumption and accuracy is completely unexplored in a noisy use-case, in which people may find themselves in crowded places with tens of other mobiles in the range of a few meters (a bus, a shopping center, or even only the queue to get into a shopping center). In those cases the loss of packets due to noise and collisions could introduce false negatives, i.e., contacts that are not traced.

Katevas et al. in a recent paper \cite{katevas2019finding} use iPhones to detect interactions between couples and groups of people. 22 people were involved in a 45 minutes experiment in which they were left in an open space and were free to socialize. The ground truth was obtained with post-processing of video recording. A total of 99 one-to-one encounters and 22 group encounters were detected. Again, the accuracy of encounter detection was satisfactory (about 89\%) but the experiment set-up does not generalize to the COVID-19 use case, for similar reasons to the previous works. The experiment was in a controlled environment, the participants had a dedicated beaconing device in one pocket and an iPhone in the other, all data were stored and post-processed with machine learning techniques. Furthermore the iPhone had access to the data produced by other sensors: acceleration, gravity, and rotation time. 

It is important to note that as of today all the experiments involving iPhones used a second dedicated device playing the role of the broadcaster. This is due to the fact that iOS does not allow an app to act as a broadcaster if the app is not in the foreground. This was noted in the literature \cite{katevas2016detecting}  and was brought to the attention of the media recently\footnote{See \url{https://www.bloomberg.com/news/articles/2020-04-20/france-says-apple-s-bluetooth-policy-is-blocking-virus-tracker}.}. This is a privacy-preserving feature that Apple introduced to prevent the exact goal of contact tracing apps: constantly tracking the user position. Even if it is likely that Apple could remove this limitation for the COVID-19 use case, there are currently no scientific works estimating the efficacy of proximity detection using an iPhone as a broadcaster for prolonged periods.

Palaghias et al. \cite{palaghias2015accurate} present an accurate technique to perform proximity detection using smartphones only. The authors start from the observation that proximity detection using RSSI only is not precise enough and elaborate a new strategy that needs 6 Bluetooth packets to correctly estimate user proximity.  Results show an accuracy close to 82\% in a realistic scenario but again the technique can not be generalized to the COVID-19 use case. Even if detection is done on-line by each phone (all other experiments performed off-line centralized detection) it uses a machine learning approach which requires training. Furthermore, proximity detection is improved by estimating the direction the user is facing, based on a previously introduced technique \cite{hoseinitabatabaei2014design} which requires access to various sensors on the phone. This technique also includes an interaction between the two phones using Bluetooth in ad hoc mode, which is different than BLE. The experimental set-up is limited to 8 people performing partly controlled operations in an indoor setting.
It is also not clear what version of Bluetooth is used in the experiment, and thus, what is the effective power consumption of the proposed technique.

\section{Discussion}

\label{sec:discussion}

The first important consideration emerging from our analysis is that, albeit called with the same name, the ``contact tracing'' needed to limit the spread of a virus is not what a mobile application can provide. A ``close contact'', according to international guidelines, is a person at a distance of less than 1-2 meters without proper protections. A smartphone app can only estimate when two devices are in communication range, regardless of where their owners are and what is in between (a thin wall, a glass\ldots), which is generally referred to as proximity detection. The second fundamental consideration is that improving contact tracing requires high precision. The reproductive number when containment measures are in place is of a few units, and the majority of the contacts are extremely predictable (family, workplace, hospital). Proximity detection should provide an estimate of, in average, less than one contact per infected person. 
The third important consideration is that a high rate of false positives could defeat the goal of the app itself (with people ignoring the messages they receive) or even be detrimental (diverting precious resources to manage false positives). 

Proximity detection, to be useful, should provide a very short list of people that had long-lasting contacts with Alice. Yet we know that Bob's phone should make a precise estimate of proximity with a very small number of samples. The literature analysis we performed shows that at the time of writing, \emph{there is no scientific evidence to support that under these conditions, a proximity detection application running on smartphones with a distributed design can provide such high precision}. 

All the works we reviewed provide reasonable accuracy for proximity detection (between 80\%-90\%) but use a set-up that is far from being applicable to the COVID-19 use case, combining at least two of the following requirements:
\begin{itemize}
	\item They require a centralized system to store all the raw data. The database is needed \emph{in order to} detect contacts and contain the RSSI for all received packets. 
	\item They require training. Calibration of the system needs ground truth provided by the experimenters, and a controlled environment.
	\item They used dedicated devices. They were not able to perform the experiments only with smartphones and used custom devices worn by participants.
	\item They required direct communications between the two phones, or access to other sources of data.
\end{itemize}

This makes impossible to forecast the accuracy of proximity detection when used ``in the wild''.

\subsection{False Positives and Sensitivity}

 Consider a person that goes to work every day with public transportation (bus), it is reasonable that he/she will stay in the bus for more than 10 minutes, and repeat this routine twice a day. If a distance of 2 m is maintained between people in the bus, and we consider only the closest 4 persons to be in communication range, we may estimate 8 contacts per day. We set the observation period to three days, that is, one day that passes from the emergence of the symptoms to when the person is tested positive (a very optimistic estimation) and two days in advance (again, a very conservative choice). This yields the number of contacts $N=24$. The numbers we introduced in \cref{sec:needle} ($R = 2$, 2 out or 6 contacts not traceable with analog contact tracing) tell that we are looking in average for 2 people every 3 infected people. If the app has full penetration in the whole society and works perfectly, then we have that every $3*24=72$ traced contacts, there are only two people that are likely positive and untracked. That is, the task of contact tracing itself, under the assumptions of a perfect application produces 35 false positives every true positive. Note that here we call a false positive someone that was in proximity of some infected person, detected by an app that has 100\% accuracy. These figures clearly state that physical closeness is not a good proxy to detect ``close contacts'' as defined in medical terms. 
Moreover, this figure does not consider several other causes of contacts (people that work in contact with the general public, social gatherings, shopping malls etc.) yet, it already shows that using an app for contact tracing may produce an unbearable amount of people to test, or people ignoring the messages they receive.



In order to make a rough estimation of the sensitivity of such an app, we rely on statistics available at the time of writing. We call $C$ the fraction of people that is successfully using the contact tracing app in a certain moment, the fraction of contacts that can be successfully detected is $S = C^2 * P$ where $C^2$ is the probability that both people in close contact run the app and $P$ is the accuracy of the app. We estimate $C$ using the following expression:

\begin{equation}
C = M * A * W
\end{equation}

Where $M$ is the fraction of people owning a smartphone, $A$ is the fraction of people willing to install the app, $W$ is the fraction of phones that supports contact tracing with BLE. 

According to the Pew Research Center\footnote{See Spring 2018 Global Attitude Survey. We restrict our analysis to advanced economies, as we focus on the European panorama:  \url{https://www.pewresearch.org/global/2019/02/05/smartphone-ownership-is-growing-rapidly-around-the-world-but-not-always-equally/}.} in advanced economies 76\% of the adult population owned a smartphone in 2018, we then set $M=0.76$. A recent survey on the acceptability of the use of a contact tracing app in several advanced economies reveals that about 74.8\% of the people would agree on installing the app, so we set $A=0.748$ \cite{altmann2020acceptability}. We note that this number is disproportionate compared to the penetration of real apps, for which numbers are extremely lower\footnote{According to the news available a the time of writing the number of downloaded apps per inhabitant is $\simeq$0.23 in Switzerland \cite{swisscovid2020monitoring}, $\simeq$0.19 in Germany \cite{koch2020report}, $\simeq$0.7 in Italy \cite{ansa2020immuni}. Note that not every downloads corresponds to a running app.}. 
Recent statistics\footnote{See the data from IDC.com including the current and past market shares of new smartphones \url{https://www.idc.com/promo/smartphone-market-share/os}.} report that Android and iOS cover almost 100\% of the market of new devices with Android alone covering 86\% of the market. 
We know that at the time of writing, Apple iOS does not allow active beaconing for applications that run in the background, and one of the work observed that 42\% of the Android smartphones did not allow beaconing even if the phone specifications theoretically allowed it \cite{girolami2018detecting}.
Assuming at some point iOS will enable beaconing on all devices we call $W_a$ the fraction of BLE-equipped Android devices that support beaconing, then if $W_a = 1 - 0.42 = 0.58$ is still a valid estimation, we have that $M = (0.14 + 0.86*W_a) \simeq 0.63$ (in Switzerland, the ratio between app downloads and active apps is 0.57 so close to our figure \cite{swisscovid2020monitoring}). This yields $C = 0.76*0.748*0.63 \simeq 0.36$. Let us assume that the app never introduces false positives (here false positive refers to the fact that the app never considers a close contact someone that is far from the subject, an extremely strong assumption given the imprecision of proximity detection) and that it is able to detect a fraction $P = 0.81$ of the proximity events, the lower bound of the results reported by the literature in non realistic conditions.
Then the sensitivity of the contact tracing application (i.e. the fraction of contacts that are correctly traced) is given by $S = C^2 * P = 0.36^2 * 0.81 \simeq 0.10$. 
That is, assuming that: i) phones are always on and carried by their owners; ii) the app is always working; iii) all the people declaring they want to use the application actually do it; iv) the precision of the app is the one observed in non-realistic controlled conditions and does not create false positives; then still the application will be able to provide \emph{around 10\% of the real contacts}.

In conclusion, in the short term, with current technologies, we argue that \textit{the high number of potential false positives and the low sensitivity does not justify the introduction of a contact tracing application with an high potential privacy risk}.




\section{The Way Ahead}
\label{sec:wayahead}

In this section we build on the previous discussion to provide some pragmatic considerations on how to make the best use of digital contact tracing.

As a first point, we stress the importance of a rigorous monitoring of the ongoing efforts to produce contact tracing applications.  It is essential that not only the source code of the apps, but also the results on their experimentation and their daily use is made available for public scrutiny. Since we outlined important technical unsolved challenges it is paramount that the way these challenges are addressed is made public, so that experts can evaluate the efficacy of the app, and validate the whole approach. It is also fundamental that the results obtained with the app will be constantly monitored during its use, to periodically asses its overall social utility.  So far we have only anecdotal evidences saying that in some countries the results of digital contact tracing are not encouraging. For instance, in Australia it was reported that even with 6 millions download, the app did not report any contact that was not already known\footnote{See \url{https://www.smh.com.au/politics/federal/much-hyped-contact-tracing-app-a-terrible-failure-20200628-p5570h.html}}, in the UK the app was suspended due to difficulties in making it working properly \footnote{See \url{https://www.technologyreview.com/2020/06/19/1004190/uk-covid-contact-tracing-app-fiasco/}.} and even in Singapore, one of the earliest adopters of digital contact tracing, where 35\% of the population appears to have downloaded the application, the government is going to provide physical tokens to citizens, to have reliable measures that are not possible with smartphones\footnote{See \url{https://www.bbc.com/news/technology-53146360}.}.

If the apps will prove to be inefficient, there are still other ways we could reshape their goals to provide some social benefit.
\subsection{From Contact Tracing to Risk Profiles}
Protection measures like safety distance, face masks and avoiding gatherings are a limitation to freedom, and certain categories of people that are less exposed to the risk may not be motivated enough to enforce them. One way to encourge them is by creating empathy towards the others, which has been shown to be be effective \cite{pfattheicher2020emotional}. To reach this goal, gamification has proven useful in several domains, including health \cite{seaborn2015gamification} and may be adopted also in this specific situation. Gamification means introducing game elements in non-gaming activities and can be used for the creation of social awareness of the consequences or one's own actions towards the more vulnerable ones, which is an effective way of motivating people to respect contention measures against COVID-19 \cite{lunn2020motivating}. 

Contact tracing apps could be reshaped to match this new purpose: Instead of trying to capture the exact number of proximity events, they could provide users with an estimation of their cumulative exposure to risk, and thus, the potential danger they represent for the others. Through gamification they could nudge users to take a better behaviour. 
The amount of information required for this task could be lower than what required for contact tracing and can be performed with anonyimzed data-sets, thus lowering the privacy risk. For instance, re-use or RPIs could make it harder to de-anonymize the identity of infected contacts, while still providing a reasonable measure of the exposure risk. Extending this concept, distributed contact tracing may support k-anonymity (as proposed recently \cite{ali2020cross}). 

\subsection{Assisted Analog Contact Tracing and the Centralized Solution}
As we mentioned, analog contact tracing requires expertise, time and can be error prone.
On the other hand we know that digital contact tracing can not be used automatically, as it only identifies proximity. Merging the two forms of tracing would instead definitely improve contact identification. If the expert can access a list of contacts which can be reviewed together with the infected person, the whole process would be made faster and more robust. 
Unfortunately this is not achievable with the distributed, privacy-preserving solution.

Instead, a centralized solution would bring a number of advantages, among which:
\begin{itemize}
	\item Proximity tracing will improve, as it could be performed with data coming from both smartphones;
	\item Big data would enable the use of advanced data analysis to filter out false positives and improve the estimation;
	\item A contact graph could be produced, which would help complex analysis and identify potential super-spreaders;
	\item If a person that tests positive is presented with a list of people (not IDs) of possible contacts, he/she may provide details on the kind of contact to reconstruct its associated risk.
\end{itemize}
Indeed, almost all the works we reviewed use a centralized approach for proximity detection, which helps overcoming several technical limitations. For instance, a centralized system that collects data for every single received BLE packet can estimate the distance from both endpoints, thus reducing the error due to multipath fading and shadowing. Yet this approach is even more privacy intrusive than what is normally described as a ``centralized system'' for contact tracing for COVID-19 \cite{troncoso2020decentralized}, which generally refers to a system that collects all the proximity events \emph{after} they were detected in the phones. Such a centralized system instead  owns enough information to de-anonymise the position of the users and their loose interactions, and in general it would be an unprecedented privacy nightmare. As scientists however, we have to ask ourselves if it makes more sense to insist on the adoption of a solution that may never work properly, or to accept the challenge of imagining technical means that may reduce the risks associated with a solution that may actually work.





\section{Conclusions}

The risk assessment of a proximity tracing app, considering its privacy issues is very hard to perform, since the fallout of data leakage can be simply impossible to predict with current information. 
For this reason, a first principle approach would call to ask what is the estimated benefit of an application that introduces a potential privacy risk.
We analyzed the available literature to answer this question, matching the data of the pandemic with the available data from experiments in proximity tracing, and our conclusion is that there is not enough evidence to support that such an app would help slow down the running contagion.

A contact tracing app, adopting the highest standards of privacy could be indeed useful to spread awareness and encourage modifications in people behavior, a goal that appears to be less daunting and more practical to achieve in the short term.

\section*{Acknowledgments}
The authors want to thank Gianrocco Lazzari for the useful feedback he provided to the realization of the paper.

\bibliographystyle{elsarticle-num-names} 
\bibliography{bibliography}

\appendix

\section{Excerpts from Relevant Documents}

\subsection{Definition of Close Contacts from National Health Institutes}
\label{app:contact}
Here we report a number of definitions of ``close contact'' from the official documents of several English speaking national health institutions. Note that most of these document extend the definition with specific provisions for, i.e., households or partners, which we don't report here as they are easy to detect without the need of a mobile phone app. The definitions may also change with time, as more evidence is accumulated on the way the virus spreads. 

The US Center for Desease and Control Prevention identifies a close contact as \emph{Individual who has had close contact ($<$ 6 feet) for a prolonged period of time} and specifies in notes that\footnote{See \url{https://www.cdc.gov/coronavirus/2019-ncov/php/public-health-recommendations.html}}:

\emph{Factors to consider when defining close contact include proximity, the duration of exposure (e.g., longer exposure time likely increases exposure risk), whether the individual has symptoms (e.g., coughing likely increases exposure risk) and whether the individual was wearing a facemask (which can efficiently block respiratory secretions from contaminating others and the environment).}

\emph{Data are insufficient to precisely define the duration of time that constitutes a prolonged exposure. Recommendations vary on the length of time of exposure from 10 minutes or more to 30 minutes or more. [\ldots] Brief interactions are less likely to result in transmission; however, symptoms and the type of interaction (e.g., did the person cough directly into the face of the individual) remain important.}

According to the Australian guidelines a close contact is a  ``\emph{face-to-face contact in any setting with a confirmed or probable case, for greater than 15 minutes cumulative over the course of a week [\ldots] sharing of a closed space with a confirmed or probable case for a prolonged period (e.g.
	more than 2 hours) in the period extending from 48 hours before onset of symptoms in the
	confirmed or probable case''}\footnote{See \url{https://www1.health.gov.au/internet/main/publishing.nsf/Content/cdna-song-novel-coronavirus.htm}}.

For the Irish Health institution a close contact is defined as \emph{Any individual who has had greater than 15 minutes face-to-face ($<$2 meters	distance) contact with a case, in any setting.}\footnote{\url{https://www.hpsc.ie/a-z/respiratory/coronavirus/novelcoronavirus/guidance/contacttracingguidance/National\%20Interim\%20Guidance\%20for\%20contact\%20tracing.pdf}}

The state of Alberta (CA) provides the following definition: \emph{individuals that lived with or otherwise had close prolonged contact (i.e., for more than 15 min and within two metres) with a case without consistent and appropriate use of PPE [Personal Protection Equipment] and not isolating}\footnote{See \url{https://open.alberta.ca/publications/coronavirus-covid-19}}
\subsection{EU Privacy Guidelines}
\label{app:docs}
The the mentioned EU Privacy Guidelines \cite{eu2020guidelines} contain the following guidelines:

\emph{The systematic and large scale monitoring of location and/or contacts between natural persons is a grave intrusion into their privacy. It can only be legitimised by relying on a voluntary adoption by the users for each of the respective purposes. This would imply, in particular, that individuals who decide not to or cannot use such applications should not suffer from any disadvantage at all}.

\emph{[\ldots] contact tracing apps do not require tracking the location of individual users. Instead, proximity data should be used; as contact tracing applications can function without direct identification of individuals,appropriate measures should be put in place to prevent re-identification; the collected information should reside on the terminal equipment of the user and only the relevant information should be collected when absolutely necessary.}

\emph{[\ldots] procedures and processes including respective algorithms implemented by the contact tracing apps should work under the strict supervision of qualified personnel in order to limit the occurrence of any false positives and negatives. In particular, the task of providing advice on next steps should not be based solely on automated processing.}

\emph{False positives will always occur to a certain degree. As the identification of an infection risk probably can have a high impact on individuals, such as remaining in self isolation until tested negative, the ability to correct data and/or subsequent analysis results is a necessity. This, of course, should only apply to scenarios and implementations where data is processed and/or stored in a way where such correction is technically feasible and where the adverse effects	mentioned above are likely to happen.}

\emph{Any server involved in the contact tracing system must only collect the contact history or the	pseudonymous identifiers of a user diagnosed as infected as the result of a proper assessment	made by health authorities and of a voluntary action of the user. Alternately, the server must	keep a list of pseudonymous identifiers of infected users or their contact history only for the time to inform potentially infected users of their exposure, and should not try to identify potentially infected users.}





\end{document}